\documentclass[prb,aps,preprint,showpacs,superscriptaddress]{revtex4-2}
\usepackage{graphicx}
\usepackage{dcolumn}   
\usepackage{bm}        
\usepackage{amssymb}
\usepackage{mathtools}
\usepackage{adjustbox}
\usepackage{lipsum}
\usepackage[english]{babel}
\usepackage{geometry}
\usepackage{booktabs}
\usepackage{array}
\usepackage{flafter}
\usepackage{float}
\usepackage{epstopdf}
\usepackage{hyperref}
\hypersetup{
	colorlinks=true,
	linkcolor=blue,
	filecolor=blue,      
	urlcolor=blue,
        citecolor=blue 
}
\begin{document}
\title{Pressure-Induced Volume Collapse and Metallization in Inverse Spinel Co$_2$TiO$_4$}
\author{Mrinmay Sahu}
\affiliation {Department of Physical Sciences, Indian Institute of Science Education and Research Kolkata, Mohanpur, Nadia, West Bengal-741246, India}
\affiliation {National Centre for High Pressure Studies, Indian Institute of Science Education and Research Kolkata, Mohanpur, Nadia, West Bengal-741246, India}
\author{Souvick Chakraborty}
\affiliation {Department of Physical Sciences, Indian Institute of Science Education and Research Kolkata, Mohanpur, Nadia, West Bengal-741246, India}
\author{Bidisha Mukherjee}
\affiliation {Department of Physical Sciences, Indian Institute of Science Education and Research Kolkata, Mohanpur, Nadia, West Bengal-741246, India}
\affiliation{National Centre for High Pressure Studies, Indian Institute of Science Education and Research Kolkata, Mohanpur, Nadia, West Bengal-741246, India}
\author{Bishnupada Ghosh}
\affiliation {Diamond Light Source Ltd, Harwell Science and Innovation Campus, Didcot OX11 0DE, United Kingdom}
\author{Asish Kumar Mishra}
\affiliation {Department of Physical Sciences, Indian Institute of Science Education and Research Kolkata, Mohanpur, Nadia, West Bengal-741246, India}
\affiliation {National Centre for High Pressure Studies, Indian Institute of Science Education and Research Kolkata, Mohanpur, Nadia, West Bengal-741246, India}
\author{Satyabrata Raj}
\affiliation {Department of Physical Sciences, Indian Institute of Science Education and Research Kolkata, Mohanpur, Nadia, West Bengal-741246, India}
\affiliation {National Centre for High Pressure Studies, Indian Institute of Science Education and Research Kolkata, Mohanpur, Nadia, West Bengal-741246, India}
\author{Goutam Dev Mukherjee}
\email [Corresponding Author:]{ goutamdev@iiserkol.ac.in}
\affiliation {Department of Physical Sciences, Indian Institute of Science Education and Research Kolkata, Mohanpur, Nadia, West Bengal-741246, India}
\affiliation {National Centre for High Pressure Studies, Indian Institute of Science Education and Research Kolkata, Mohanpur, Nadia, West Bengal-741246, India}
\date{\today}
\begin{abstract}
The structural, vibrational, electronic, and magnetic properties of inverse spinel $Co_2TiO_4$ (CTO-Sp) under high-pressure (HP) conditions are systematically investigated using X-ray diffraction, Raman spectroscopy, in situ optical microscopy, and first-principles density functional theory (DFT) calculations. At ambient conditions, CTO-Sp exhibits a cubic phase with a space group $Fd\bar{3}m$, and it undergoes two notable structural phase transitions at HP. The first transition, occurring at approximately 7.3 GPa, leads to the tetragonal-$I4_1/amd$ phase with minimal alteration in unit cell volume. {The second transition takes place near 17.3 GPa, where two orthorhombic phases emerge and coexist above this pressure.} This second structural transition corresponds to a first-order phase transition involving a significant reduction in unit cell volume of approximately 17.5$\%$. The bulk compressibility of CTO-Sp and its HP post-spinel phases is almost equal to the average polyhedral compressibility within each phase.  DFT calculations reveal a high-spin to low-spin transition, accompanied by the collapse of local magnetic moments in the $Cmcm$ orthorhombic phase, leading to the sample's pressure-induced metallization.                    
\end{abstract}
\maketitle
\section{Introduction}
Binary and ternary spinels have attracted significant attention, not solely owing to their distinctive physical properties, including multiferroicity, unconventional magnetic behaviour, negative thermal expansion, and magnetodielectric effects, but also due to their diverse range of applications in high-frequency electronic components, sensors, and ultrahigh-density magnetic recording media \cite{Tomiyasu2004,Yamasaki2006,Rossi2019,Choi2009,Suzuki2008,Camley2009,Naito1971,Harris2009}. 
Spinel oxides follow the general formula $AB_2O_4$, where A and B represent transition metal ions. Based on the distribution of cations, spinels are categorized into two types: normal spinel, characterized by the formula [$A^{2+}]_T[(B^{3+})_2]_oO_4$], and inverse spinel, identified by the formula [$B^{3+}]_T[(A^{2+}B^{3+})]_oO_4$]. In the normal spinel, $[A^{2+}]_T$ occupies the tetrahedral site, while two $[B^{3+}]_o$ cations occupy octahedral sites. This arrangement differs from the inverse spinel, where $[B^{3+}]_T$ cations are in tetrahedral sites, and $[A^{2+}B^{3+}]_o$ cations are in octahedral sites within a single formula unit. In an inverse spinel structure, 50\% of the octahedral sites are occupied by A cations, while B cations are distributed between both octahedral and tetrahedral sites \cite{Millard1995}.
At ambient conditions, the majority of spinel cobaltites and ferrites crystallize into normal cubic symmetry, but this configuration becomes unstable under high-pressure (HP), as well as at elevated temperatures \cite{Ghosh2018,Rahman2017}. 
Increasing pressure or temperature induces disorder in the tetrahedral [A] and octahedral [B] sites, subsequently adjusting the Curie temperature of the magnetic spinel and triggering a magnetic transition \cite{Boyanov1994,Subias2013,Rozenberg2007}. 
Normal and inverse spinels exhibit a range of structural phase transitions under pressure, leading to the formation of various polymorphs. Within their pressure/temperature-induced post-spinel phases and diverse polymorphic forms, numerous intriguing properties have been discovered. These include electronic characteristics such as insulator-to-metal transitions (IMTs), superconductivity, spin reorientation linked to magnetic transitions, Mott transitions, high spin to low spin transitions, and colossal magnetoresistance. These phenomena arise from the delicate interplay between crystal lattice distance and electronic degrees of freedom. The thermodynamic properties of numerous spinels can also be externally adjusted through perturbations, such as changes in pressure and temperature \cite{Rahman2017,Suzuki1999,Arielly2011, Javier2014}. 

In the field of Earth sciences, there has been significant interest in the pressure-induced phase transformations of titanate spinels ($A_2TiO_4$-Sp; where A can be any transition metal). These spinels have captured attention due to their conventional classification as low-pressure analogs of ringwoodite, the most abundant mineral in the lower section of the mantle transition zone. Therefore, the exploration of the HP behaviour of the inverse spinel, $Co_2TiO_4$ (CTO-Sp), is anticipated to provide crucial insights into understanding the dynamics of Earth's lower mantle \cite{Senz2001,Millard1995,Shim2001,Oneill1984,Moorbath1969}. 
Cobalt orthotitanate, CTO-Sp, typically adopts the cubic crystal structure with the space group $Fd\bar{3}m$. Its electronic configuration is commonly assumed to be $(Co^{2+})_T (Co^{2+}Ti^{4+})_OO_4$, where the subscript T and O represent the tetrahedral and octahedral sites, respectively \cite{Nayak2015}. 
{The findings show that CTO-Sp is a rare material with unusual magnetic properties. It exhibits a semi-spin glass (SSG) behaviour. This follows a quasi-long-range ferrimagnetic ordering transition at $T_C = 48 K$. Additionally, magnetic compensation occurs around $T_{comp} = 32 K$, where the magnetizations of the two sublattices cancel each other.} Nevertheless, there remains a dispute regarding the existence of SSG state in CTO-Sp \cite{Nayak2015,Wei2017,Nayak2016,NayakCo2016}. 
Random anisotropy is a key factor influencing the global behaviour of CTO-Sp. This anisotropy arises from unsystematic lattice distortion, which screens the local charge density fluctuation caused by the substantial charge difference between transition metal ions \cite{Gavoille1991}.   
Pressure can induce lattice distortion and modify the shielding of local charge density fluctuations, resulting in a modification of the global behaviour of CTO-Sp.  
Numerous experimental investigations utilizing various techniques have been undertaken to observe the HP post-spinel phase with varying polymorphs and to establish the equation of state for $A_2TiO_4$ spinels \cite{Wang2002, Senz2001,Millard1995}. 
{For instance, Wang et al. reported that spinel $Zn_2TiO_4$ begins to transform into an orthorhombic crystal structure at 23.7 GPa, with the phase transition completing at 32.4 GPa.\cite{Wang2002}. A Jahn-Teller transition from a cubic to a tetragonal structure at approximately 9 GPa has been documented in $Fe_2TiO_4$. Additionally, an HP transition from the tetragonal to the orthorhombic structure of $Fe_2TiO_4$ has been identified at around 11 GPa, along with the discovery of a higher-pressure polymorph at approximately 45 GPa \cite{Yamanaka2009}}. 
Theoretical simulations by Zhang et al.\cite{Zhang2019} indicate that CTO-Sp exhibits several post-spinel phases with various stable polymorphs in the HP range. They also predict a potential pressure-induced magnetic transition in the HP post-spinel phase.\\ 

{Unfortunately, there is a lack of high-pressure data to validate the theoretical predictions regarding CTO-Sp. Therefore, the primary objective of this study is to investigate CTO-Sp under HP conditions to understand its post-spinel polymorphism, along with a potential pressure-induced magnetic transition associated with a Mott insulator-to-metal transition (MIT)\cite{Zhang2019}. We synthesized polycrystalline CTO-Sp and examined its structural, vibrational, and electronic properties under pressure. A possible magnetic transition was also investigated. This study combines X-ray diffraction (XRD), Raman spectroscopy, and first-principles density functional theory (DFT) calculations. In contrast to the findings of Zhang et al.\cite{Zhang2019}, who reported the stability of CTO-Sp up to ~21 GPa, our results reveal two distinct structural phase transitions below this pressure: a cubic-to-tetragonal transition near 7.3 GPa, followed by a tetragonal-to-orthorhombic transition around 17.3 GPa. Notably, the latter transition is accompanied by a significant volume collapse across the phase boundary.}

\section{Experimental Details}
The inverse spinel $Co_2TiO_4$ polycrystalline sample was synthesized via the conventional solid-state reaction route \cite{Liu2021,Hubsch1982}. A stoichiometric mixture of $Co_3O_4$ (99.9 \% purity) and $TiO_2$ (99.5 \% purity) powders, obtained from Sigma-Aldrich, was thoroughly blended using an agate mortar and pestle. The resulting mixture underwent calcination at 1100 $^{\circ}$C for 24 hours to produce the pure CTO-Sp. The calcined powder was then compressed into pellet form under a hydrostatic pressure of 2.2 kbar, followed by sintering at 1200 $^{\circ}$C for 20 hours in an air environment to achieve a high-density polycrystalline CTO-Sp sample. {Subsequently, the sintered CTO-Sp pellet was crushed and ground for a long time to obtain a fine powder, which was used for all experimental purposes.} The phase purity of the polycrystalline CTO-Sp was verified using room temperature ambient pressure XRD measurements \cite{Liu2021}. Energy Dispersive X-ray Spectroscopy (EDX) measurements were conducted to determine the average stoichiometric ratio of the sample. The grain size and distribution of the CTO-Sp powder were examined using Field Emission Scanning Electron Microscopy (FESEM). Detailed discussions on EDX data and FESEM data can be found in the supplemental information.
We conducted room temperature ambient and HP x-ray diffraction measurements for CTO-Sp at the XPRESS beamline in the ELETTRA synchrotron radiation facility in Trieste, Italy, utilizing X-ray beam at the wavelength of 0.4957 $\AA$. All HP experiments were carried out using a piston-cylinder type diamond anvil cell (DAC) from Almax easy Lab, with 300 $\mu$m dia diamond culets. A 290 $\mu$m thick steel gasket was pre-indented to a thickness of 40 $\mu$m using the DAC. This pre-indented steel gasket with a central hole of diameter 100 $\mu$m, was placed between two oppositely faced diamond anvils to create a suitable sample chamber. A minute amount of powder sample, approximately 50 $\mu$m in diameter, was loaded inside the sample chamber. Additionally, a small quantity of silver powder was mixed with the sample to serve as a pressure calibrant for HP XRD measurements. The 4:1 methanol-ethanol mixture was employed as a liquid pressure transmitting medium (PTM). The incident X-ray beam was collimated to a spot size diameter of 50 $\mu$m and directed through the center of the gasket hole inside the DAC. The diffraction data were recorded using a PILATUS 3S 6M detector positioned perpendicular to the incident beam. The sample-to-detector distance was determined using the XRD patterns of the standard sample $CeO_2$. DIOPTAS software was utilized to convert all 2D XRD images into $2\theta$ vs intensity profiles \cite{prescher2015dioptas} 
The XRD patterns were indexed using CRYSFIRE software \cite{Shirley1999}. GSAS software was employed to conduct Lebail and Rietveld refinements of the XRD patterns \cite{toby2001expgui}.
Raman spectra, both ambient and pressure-dependent, were acquired in the backscattering geometry using Monovista from SI GmbH, a confocal micro-Raman system consisting of a 750 mm monochromator and a back-illuminated PIXIS 100BR (1340×100) CCD camera. The sample was excited with the 532 nm laser from the Cobalt-samba diode pump laser source. To prevent local heating, the laser power was maintained at a constant level of approximately 20 mW. A 20$X$ objective lens with infinite correction, offering a large working distance and a laser spot diameter of about 5 $\mu$m, was utilized for collecting the scattered radiation. A grating featuring 1500 grooves/mm was used to disperse the scattered light, achieving a spectral resolution better than 1.2 $cm^{-1}$. Throughout the entire experiment, the scattered signal collection area of the sample remained fixed. Ruby chips (approximate size 5-8 $\mu$m) were loaded alongside the powder sample in the sample chamber, serving as pressure markers during Raman spectroscopy measurements \cite{Mao1986}. The ruby chips were positioned close to the sample. The sample loading technique inside the DAC was consistent with the description provided earlier. Similar to XRD, methanol-ethanol was utilized as a liquid PTM during HP Raman spectroscopy measurements. All experiments were conducted using the polycrystalline powders of $Co_2TiO_4$ sample.

\section{Computational Details}
Density Functional calculations were carried out using the Vienna Ab Initio Simulation Package (VASP) \cite{kresse1996efficient} with the generalized gradient approximation exchange-correlation functional in the form of the Perdew-Burke-Ernzerhof (GGA-PBE) \cite{perdew1996generalized}. The frozen-core Projector Augmented Wave (PAW) \cite{blochl1994projector} technique with an energy cutoff of 500 eV for the plane wave basis was employed for the calculations. {The initial lattice constants and the atomic positions were taken from the experimental data. Afterwards, during structural relaxation, the atomic positions were allowed to relax while keeping the lattice constants fixed.} Each of the unit cells for the different phases contains a different number of formula units and different lattice constants, and as such, the $k$-point grid is chosen in such a way that the product of the lattice constant and the number of $k$-points along any given direction is $\sim$60. For all the calculations, the energy convergence tolerance was set at $10^{-8}$ eV and the force tolerance for ionic optimization was kept at $10^{-3}$ eV/Å. The GGA+U method is used to treat the partially filled $d$ orbitals of the Co atoms with U$_{eff}$ = 4.0 eV kept fixed for all the phases. {The criterion for selecting U$_{eff}$ = 4.0 eV is discussed in the supplemental information.}   

\section{Results and Discussion}
\subsection{High-pressure x-ray powder diffraction}
We have conducted HP X-ray powder diffraction measurements on CTO-Sp, covering the pressure range from ambient conditions to 25.6 GPa, as illustrated in Figure 1. The reflection lines in the ambient XRD pattern were successfully indexed with the cubic phase, possessing the space group Fd$\bar{3}$m. The Rietveld analysis of the ambient XRPD pattern, depicted in Figure 2(a), confirms the stabilization of the cubic phase. Detailed information on the refined structural parameters is provided in Table I. The calculated unit cell parameters for the best fit are a = b = c = 8.4473(7)$\AA$ with the corresponding unit cell volume, $V_0$ = 602.788(13) $\AA{^3}$. Our data exhibits excellent agreement with the reported literature \cite{Liu2021}. In the cubic phase, up to approximately 7.2 GPa, all the Bragg peaks are very sharp and well indexed with the space group Fd$\bar{3}$m. However, at and above 8.7 GPa, we observed broadening in some reflection lines. For our HP XRD measurements, we utilized a methanol-ethanol (4:1) mixture as a pressure-transmitting medium, which maintains hydrostatic conditions within the sample chamber up to 10 GPa \cite{Klotz2009,Chen2021}. Therefore, the broadening of the reflection lines can not be attributed to the presence of any pressure gradient inside the DAC. The Bragg reflection line attributed to the Ag pressure marker remained sharp with a FWHM value of approximately  0.030(5)$^{\circ}$ up to about 9.6 GPa. A careful inspection of the XRD pattern at 9.6 GPa showed splitting of the (4,0,0) Bragg peak of the cubic phase (Supplementary Fig.S3). A re-indexing of the XRD pattern at 9.6 GPa resulted in a tetragonal structure with corresponding $I{4_1}/amd$ crystal symmetry.
The Rietveld analysis of the XRD pattern of CTO-Sp at 9.6 GPa, shown in Figure 2(b), indicates the stabilization of the tetragonal phase. The cell parameters for the best fit at 9.6 GPa are a = b = 5.8845(7) $\AA$ and c = 8.3374(5) $\AA$ with the unit cell volume, $V_0$ = 288.711(1) $\AA{^3}$. The refined lattice parameters are very close to the unit cell parameters of the single-crystal $Fe_2TiO_4$ \cite{Yamanaka2009}. Detailed information regarding the refined structural parameters is provided in Table I. A Rietveld refinement of the XRD pattern at 8.7 GPa in the new tetragonal phase produced a better fit. In the cubic phase, the lattice parameter and unit cell volume exhibit a linear decrease with applied pressure up to approximately 7.3 GPa, as depicted in Figure 3(a) and Figure 4(a), respectively. Following the cubic to tetragonal phase transition around 8.7 GPa, a reduction in the lattice parameter "a" and an increase along the c-direction are observed (see Figure 3(a)). The tetragonality (c/a) of the unit cell in the tetragonal phase rises with pressure, reaching its maximum at 16.1 GPa, as illustrated in the inset of Figure 3(a). This increase in tetragonality typically signifies a transition to a lower symmetry structure. In Figure 1, at 17.3 GPa, a new diffraction peak at $2\theta = {10.7}^{\circ}$ is observed, and another peak at $2\theta = 13.7^{\circ}$ begins to split at the same pressure. The intensity of the newly observed peak increases with pressure, reaching its maximum at the highest pressure of this study. At 19 GPa, a new broad peak around $2\theta = {16}^{\circ}$ is observed, along with a peak splitting at $2\theta = {17.8}^{\circ}$. Beyond 19 GPa, additional new diffraction peaks and peak splitting are observed (refer to Figure 1, where newly identified peaks are denoted by asterisks). All these newly observed peaks become more prominent under pressure, and the split peaks become more distinguishable with increasing pressure. {The observation of new intense peaks and the splitting of diffraction peaks indicate a possible structural phase transition of CTO-Sp above 17 GPa. Notably, the XRD patterns are best indexed to two distinct orthorhombic phases with space groups Fddd and Cmcm. In Figure 2(c), we have shown the Rietveld refinement analysis of the diffraction data at 19.1 GPa, which shows the excellent fit to space groups Fddd and Cmcm.} The determined lattice parameters for the optimal fit of the XRD pattern at 19.1 GPa are as follows: (i) Fddd-orthorhombic: a = 8.2650(2) $\AA$, b = 8.3321(7) $\AA$, c = 8.1670(6) $\AA$, and $V_0$ = 562.41(6) $\AA^{3}$. (ii) Cmcm-orthorhombic: a = 2.7521(7), b = 8.8650(2), c = 9.5185(8), and $V_0$ = 232.220(4)$\AA^{3}$. Detailed information regarding the refined structural parameters is presented in Table II. However, the phase fraction percentages for the Fddd and Cmcm space groups are 85.8 $\%$ and 14.2$\%$, respectively, at 17.3 GPa.

{Zhang et al. \cite{Zhang2019} did not observe any structural phase transition below 21 GPa in their high-pressure XRD measurements. They reported the possibility of a structural phase transition occurring above 21 GPa, based on the emergence of new Bragg peaks. However, their study lacks sufficient experimental data to clearly identify or characterize the post-spinel phases. In particular, only a limited number of data points were collected beyond the proposed transition pressure, making it difficult to draw definitive conclusions about the structural changes. Furthermore, they did not carry out Rietveld refinement on their XRD patterns, which limits the reliability of phase identification and the determination of precise crystallographic information. This may explain why the intermediate phase transitions observed in our study were not detected in theirs.}

To assess the stability of the four distinct phases under different pressures - cubic (Fd$\bar{3}$m), tetragonal (I{$4_1$}/amd), and two orthorhombic phases (Fddd and Cmcm) - first principle calculations were performed using the DFT + U formalism, with $U_{eff}$ = 4.0 eV for the Co atoms. The optimized crystal structures for these phases are depicted in Figure 5. It is interesting to see the unit cell arrangement of inverse spinel CTO. Typically, in normal spinels with general formula $AB_2O_4$, $A^{2+}$ cations reside on the tetrahedral site (A site), while $B^{3+}$ cations occupy the octahedral site (B site). However, in $Co_2TiO_4$, cobalt exists single valence states, $Co^{2+}$ \cite{Ghosh2018}. {Our Bader charge analysis of Co valency at different sites in $Co_2TiO_4$ confirms that the Co atoms exhibit a +2 oxidation state. Further details are provided in the Supplemental Information.} Consequently, $Co^{2+}$ occupies both the A and B sites in equal occupations, while the remaining $50\%$ of the B site is filled by the $Ti^{4+}$ ions. Hence, the formula for the inverse spinel CTO can be expressed as $Co(Co_{0.5}Ti_{0.5})_2O_4$. In the cubic, tetragonal, and Fddd-orthorhombic phases, the unit cell of CTO-Sp is formed by a combination of $Co^{2+}O_4$ tetrahedra and $Co^{2+}/Ti^{4+}O_6$ octahedra. Conversely, in the case of Cmcm space group, the unit cell is formed by a herringbone octahedral array. In this array, $Co^{2+}$ occupies A site and $Co^{2+}/Ti^{4+}$ are disorderly situated at B site. The arrangement of the cation-oxygen polyhedra for each phase is shown in Figure 5. The pressure-dependent phase fraction for the two different space groups in the orthorhombic phase is illustrated in Figure 3(b). Notably, the phase fraction for the Cmcm space group increases with pressure, while the phase fraction for the Fddd-orthorhombic phase decreases with pressure. At 17.3 GPa, the new diffraction peak at $2\theta = {10.7}^{\circ}$ is responsible for the presence of the Cmcm space group. The intensity of this new diffraction peak increases with rising pressure, directly correlating with the increase in the Cmcm-phase fraction percentage. Conversely, all peaks responsible for the Fddd space group gradually broaden under high pressure, which can be attributed to the decrease in the phase fraction percentage. At the highest pressure of around 25 GPa, the phase fraction for the Fddd and Cmcm space groups is approximately 19 $\%$ and 81 $\%$, respectively, as indicated in Figure 3(b). In the orthorhombic phase, within the Fddd space group, the c-axis displays greater compressibility compared to the other two axes. Conversely, in the Cmcm space group, the compressibility of the b-axis is notably higher than that of the other two axes, see Figure 3(a). This directional variation in compression, known as anisotropic compression, implies that the bulk modulus can vary along distinct crystallographic directions in the post-spinel phase of CTO-Sp. Above 25 GPa, Rietveld refinement becomes challenging as the diffraction peak broadens with reduced intensity, hindering accurate analysis.                

For the calculation of the isothermal bulk modulus ($B_0$) of CTO-Sp, we applied a least-squares method to fit the P-V data obtained from ambient pressure to 25 GPa using the third-order Birch-Murnaghan equation of state (BM-EoS) \cite{birch1947finite,angel2014eosfit7c}:
\begin{equation}
	P(V)=\frac{3 B_{0}}{2}\left[\left(\frac{V_{0}}{V}\right)^{7 / 3}-\left(\frac{V_{0}}{V}\right)^{5 / 3}\right] \times\left\{1+\frac{3}{4}\left(B_{0}^{\prime}-4\right)\left[\left(\frac{V_{0}}{V}\right)^{2 / 3}-1\right]\right\}
\end{equation}
where $B_0'$ is the first-order pressure derivative of $B_0$, $V_0$  is the unit cell volume at ambient pressure, and $V$ is the unit cell volume at any pressure.
The data fitted to the model is illustrated in Figure 4(a), while the obtained EoS parameters for various phases are detailed in Table III. Specifically, the bulk modulus ($B_0$) for the cubic phase is determined as 203(2) GPa, and for the tetragonal phase, it is 235(7) GPa. In the orthorhombic phase with the Fddd space group, the bulk modulus is 217(2) GPa, and for the Cmcm, it is 230(2) GPa. The obtained bulk modulus of CTO-Sp in both the spinel and post-spinel phases is comparable to that of similar oxide spinels in their ambient and high-pressure polymorphs \cite{Recio2001, Asbrink1998}. {Zhang et al. \cite{Zhang2019} reported a bulk modulus of about 175 GPa for CTO in the cubic phase. Our measured value is slightly higher. We have carried out Rietveld refinement at every pressure point to accurately estimate the structural parameters. Furthermore, we have carried out the EoS fit of the parent cubic phase till about 7.3 GPa. This possibly have resulted in a slightly different value of bulk modulus. }

{Our results show similar or slightly higher incompressibility in the high-pressure orthorhombic phases. The bulk modulus is sensitive to structural changes and volume. At the tetragonal-to-orthorhombic transition, a sudden volume collapse occurs, often causing atomic rearrangements that can reduce the bulk modulus. This behavior is common in several other materials. For example, in $Ba_{0.5}Sr_{0.5}CuSi_4O_{10}$ and clinochlore, the bulk modulus dropped significantly after structural transitions due to bond relaxation \cite{Knight2013, Soldavini2024}. In $CoO$, a cubic-to-rhombohedral transition also led to a slight modulus decrease \cite{Guo2002}. These examples highlight that a drop in bulk modulus can occur with volume collapse if the new structure is more flexible or has weaker bonding.}  

We will now focus on the bulk moduli of various octahedra within different phases of CTO-Sp. Using a third-order Birch-Murnaghan equation of state (BM-EoS), we have fitted the pressure-dependent change in polyhedral volume data, ranging from ambient pressure to 25 GPa. This approach allows us to obtain the polyhedral bulk moduli for different phases, and the corresponding fitted data are depicted in Figures 4(b) and (c). The values for the polyhedral bulk moduli and their pressure derivatives are summarized in Table III. In the cubic phase, the tetrahedral bulk modulus $(B_{0})_{tet}$ and the octahedral bulk modulus $(B_{0})_{oct}$ are determined as 204(2) GPa and 202(1) GPa, respectively. It is noteworthy that the compressibility of the two distinct octahedra in the cubic phase is almost identical. The average value of these polyhedral bulk moduli ($\overline{B_0}$ = 203(3) GPa) is the same as the macroscopic bulk modulus ($B_0 = 203(2)$)of CTO-Sp in the cubic phase (refer to Table III). Moving to the tetragonal phase, the tetrahedral and octahedral bulk moduli are found to be 260(1) GPa and 236(7) GPa, respectively. Remarkably, the tetrahedra in the tetragonal phase exhibit lower compressibility compared to the octahedra. The compression and distortion of the tetrahedral unit cell are predominantly achieved through the deformation of the octahedra within the unit cell. The average bulk modulus of these two octahedra ($\overline{B_0}$) is 248(8) GPa, nearly in line (within the error limits) with the bulk modulus ($B_0$ = 235(7) GPa) of CTO-Sp in the tetragonal phase. For the orthorhombic-Fddd space group, the BM-EoS fit reveals a tetrahedral bulk modulus of approximately 222(1) GPa and an octahedral bulk modulus of around 206(4) GPa. Once again, the octahedra in the Fddd-orthorhombic phase exhibit greater compressibility compared to the tetrahedra. In the case of the Cmcm space group, the structure comprises a herringbone octahedra array, as discussed previously. The bulk modulus of $Co^{2+}O_6$ octahedra is estimated to be around 232(2) GPa, while for the $Co^{2+}/Ti^{4+}O_6$ octahedra, it is approximately 231(4) GPa. For both the orthorhombic phases, the average polyhedral bulk modulus is comparable with the macroscopic bulk modulus of the respective space groups (see Table III). Therefore, the bulk compressibility of inverse spinel $Co_2TiO_4$ and its HP post-spinel phases can be estimated as the average polyhedral compressibility within each phase. Recio et al. observed that the bulk compressibility in the direct and inverse spinel-type compounds can be expressed by considering cation oxide polyhedral compressibility along with a term accounting for the hydrostatic pressure effect on the oxygen position in the unit cell \cite{Recio2001}. Their theoretical studies indicate that, for many spinel-type compounds, the macroscopic bulk modulus can be estimated as the average of polyhedral bulk moduli \cite{Recio2001}. 

The variability in compressibility across different polyhedra prompts an investigation into the pressure-dependent changes in bond angle variation and distortion indices (bond length) for various polyhedra. The bond angle variance ($\sigma^2$) and the distortion index ($DI$) of the polyhedra can be defined as follows \cite{Vesta2019}:
\begin{equation}
	\sigma^2 = 1/(N-1)\sum_{i=1}^{N}{(\theta_i-\theta_0)^2}
\end{equation}
Here, $N$ represents the number of bond angles in the polyhedron, $\theta_i$ is the $i$-th bond angle, and $\theta_0$ is the ideal bond angle for a regular polyhedron.
\begin{equation}
	DI = 1/N\sum_{i=1}^{N}{(|l_i-l_{av}|)/l_{av}}
\end{equation}
In this equation, $l_i$ denotes the distance from the central ion to the $i$-th coordinating ion, and $l_{av}$ is the average bond length.

In Figure 6(a), it is evident that the tetrahedral bond angle variance is zero in the cubic phase and begins to increase in the tetragonal phase. This increase in bond angle variance is directly associated with the tetrahedral distortion within the system. The escalating bond angle variance signifies a heightened tetrahedral distortion attributed to the Jahn-Teller effect. Thus, the transition from cubic to tetragonal phase is instigated by the tetrahedral distortion due to the Jahn-Teller effect of $Co^{2+}$ cation at the tetrahedral site. Concerning the octahedral distortion, the index remains zero up to around 7.2 GPa. Beyond this pressure, in the tetragonal phase, it reaches a maximum value of $9.1 \times 10^{-4}$, followed by a minimum at about 10.4 GPa, as depicted in Figure 6(b). Above 10.4 GPa, the octahedral distortion index sharply increases, leading to a tetragonal to orthorhombic structural phase transition at around 17.3 GPa. The deformation of the octahedra within the tetragonal unit cell is responsible for the increase in tetragonality and distortion of the unit cell. In the orthorhombic phase within the Fddd space group, the distortion index of the octahedra decreases with pressure up to 20.2 GPa, followed by a rapid increase beyond this pressure. This increase in octahedral distortion indicates instability in the Fddd-orthorhombic phase at HP. For the octahedra in the Cmcm-orthorhombic phase, the distortion index of the $Co^{2+}O_6$ octahedra increases up to 20.2 GPa, followed by saturation beyond that pressure, as shown in Figure 6(c). The distortion index of $Co^{2+}/Ti^{4+}O_6$ octahedra in the Cmcm-orthorhombic phase decreases with pressure up to the highest pressure value. The saturation of the distortion index for $Co^{2+}O_6$ octahedra and a decreasing trend in the distortion index for $Co^{2+}/Ti^{4+}O_6$ octahedra imply a rise in the phase fraction percentage along with the stability of the Cmcm-orthorhombic phase at the higher pressures.

\subsection{High-pressure Raman study}
As a complementary study to XRD and also to probe the phonon dynamics and the pressure effect on phonon modes, we have performed high-pressure Raman spectroscopy measurements at ambient temperatures on the inverse spinel polycrystalline $Co_2TiO_4$. In the literature, as far as our knowledge goes, there have not been any systematic HP Raman measurements on CTO-Sp. At ambient conditions, CTO-Sp crystallizes in cubic phase with the space group $Fd\bar{3}m$ (no. 227, Z = 8). Hence, according to the group theory analysis, the vibrational modes of CTO-Sp at $\Gamma$ point are as follows:
$\Gamma$ = $A_{1g}$ + $E_g$ + $3A_{2u}$ + $3E_u$ + $3T_{2u}$ + $3T_{2g}$ + $7T_{1u}$ + $T_{1g}$. Among which five modes, ($A_{1g}$ + $E_g$ + $3T_{2g}$) are Raman active. At ambient, we have observed four phonon modes in the Raman spectrum of CTO-Sp, as indicated in Figure 7(a). Two weak phonon modes centered at 174 $cm^{-1}$ and 309.4 $cm^{-1}$ can be identified as $T_{2g}(1)$ and $E_g$ modes respectively. The translational movement of the entire $CoO_4$ tetrahedron accounts for the mode centred at 174 $cm^{-1}$ ($T_{2g}(1)$). The phonon mode $E_g$ at 309.4 $cm^{-1}$ corresponds to the symmetric bending of oxygen ions with respect to cations in a tetrahedral ($CoO_4$) environment. The asymmetric bending and stretching of oxygen anion in the $Co/TiO_6$ octahedra cause a broad peak centered at 531.6 $cm^{-1}$ ($T_2g(2)$). The robust and intense peak centered at 701.8 $cm^{-1}$ corresponds to the symmetric stretching of oxygen atoms in the $CoO_4$ tetrahedra and can be considered $A_{1g}$ symmetry. This $A_{1g}$ mode is often referred to as the tetrahedral breathing mode of spinels. These Raman modes are all assigned following the reported literature by Prosnikov et al. and others \cite{Prosnikov2016,Wang2020,NayakCo3O42016}. 

We will now examine the impact of external pressure on the phonon modes of polycrystalline CTO-Sp. Figure 7(b) illustrates the pressure-dependent Raman spectra at selected pressure values. All Raman modes shift to higher energies with increasing pressure. {Under high pressure, Raman modes typically shift to higher frequencies (blue shift) due to increased interatomic force constants. As atoms are compressed, bond lengths decrease, making the bonds stiffer and increasing the restoring forces. According to the creation, the phonon frequency, $\omega = \sqrt{\frac{k}{\mu}}$, where $k$ is the force constant and $\mu$ is the reduced mass. An increase in $k$ increases $\omega$. However, some modes may shift lower (red shift) due to structural instability, anharmonicity, or phase transitions. Overall, blue shifts indicate stronger bonds and lattice compression.} 
{From Figure 7(b), it is evident that the $T_{2g}(1)$ mode disappears above 5.7 GPa. The $E_g$ mode progressively broadens with increasing pressure and vanishes beyond 12.5 GPa. Around 18.1 GPa, a new peak, labeled M1, emerges at 690 $cm^{-1}$}. To thoroughly examine the Raman shifts and the variation in linewidth of different modes under pressure, each peak in the Raman spectrum is modeled using a Lorentzian profile, which is well-suited for lifetime-broadened peaks. Baseline correction is not applied to the spectra before fitting, as there are no detectable contributions from fluorescence or continuum signals. The pressure-dependent Raman mode frequencies are plotted in Figure 8(a). The $T_{2g}(1)$ mode, initially centered at 174 $cm^{-1}$ under ambient conditions, exhibits a linear pressure dependency with significant broadening. The $E_g$ mode undergoes a slope change around 7.3 GPa, becoming broader with pressure and disappearing beyond 12.5 GPa. This behaviour correlates with the structural phase transition from cubic to tetragonal phases, driven by tetrahedral distortion due to the Jahn-Teller effect, as discussed in the HP XRD results. The mode originating from the $Co/TiO_6$ octahedra, $T_{2g}(2)$, exhibits a slope change above 17.3 GPa, while the tetrahedral breathing mode, $A_{1g}$, shows a slight slope at the same pressure. The change in slope of $T_{2g}(2)$ and $A_{1g}$ modes and the emergence of the new peak M1 between 17-18 GPa indicate a structural reorientation of the unit cell of polycrystalline CTO-Sp, corroborated by earlier XRD measurements. A structural transition from tetragonal to orthorhombic phase was observed around 17.3 GPa. {The newly observed mode M1 most likely originates from the orthorhombic-Cmcm phase. Moreno et al. \cite{Moreno2011} reported ab initio phonon calculations for $ZnGa_2O_4$ (structurally similar to CTO) in the high-pressure Cmcm phase, assigning a Raman mode at 692 $cm^{-1}$ as $A_g$. Based on this, we propose that the mode M1 around 690 $cm^{-1}$ in high-pressure CTO is likely the $A_g$ mode of the orthorhombic-Cmcm structure.} The presence of $A_1g$ tetrahedral breathing mode at higher pressures with minimal slope change across the tetragonal to orthorhombic phase transition indicates a zero distortion index associated with the cation-oxygen bond lengths, as mentioned in the HP XRD study. A triply degenerate mode cannot exist in a tetragonal or orthorhombic structure; it should split. But we observed no splitting of the $T_{2g}(2)$ mode in the tetragonal phase, which could be due to several reasons. The splitting may be very small due to subtle symmetry-lowering effects in the tetragonal phase, making it difficult to resolve experimentally. Additionally, the $T_{2g}(2)$ mode is not sharp, even under ambient conditions.	The dynamic disorder could average out local distortions, effectively restoring higher symmetry on the timescale of Raman scattering, thus preventing the expected splitting of the modes. Strong electron-phonon interactions could lead to a renormalization of vibrational modes, masking the splitting that would otherwise occur. Such interactions may be prominent in CTO due to the presence of transition metal ions with partially filled d-orbitals. The $T_2g(1)$ mode, which has very low intensity after loading the sample into the diamond anvil cell, makes it challenging to track the splitting. We also refer to studies by Kyono et al. \cite{kyono2011} on $Fe_2TiO_4$ and Wang et al. \cite{Wang2020} on $Mg_{2}TiO_4$, which report similar behaviour, where the $T_{2g}$ Raman modes do not show clear splitting in the high-pressure tetragonal and orthorhombic phases, although broadening is observed. {The ambient Raman data of the decompressed sample, shown in Figure S4 of the supplemental information, indicates that the phase transition is reversible.} Interestingly, the full width at half maximum (FWHM) of the $T_{2g}(2)$ and $A_{1g}$ modes exhibits a minimum at approximately 17.3 GPa, as illustrated in Figure 8(b). The FWHM of the Raman mode is inversely related to the lifetime of the phonon mode. Increasing pressure gradually introduces strain within the lattice, expected to lead to a reduction in the lifetime of phonons. However, the decrease in phonon-mode FWHM indicates a decrease in anharmonic phonon interactions. A minimum in FWHM of phonon mode suggests a possibility of strong electron-phonon interactions, which causes a decrease in the lifetime of phonons after an electronic transition \cite{Gupta2017,Bera2013,Saha2018}. The Gruneisen parameter is widely recognized for its pivotal role in understanding heat capacities and vibrational entropies using the Kieffer model \cite{Kieffer1979,Kieffer1979_1}. By employing the bulk modulus derived from our HP XRD data and using the pressure derivative of a vibrational frequency, the Gruneisen parameter ($\gamma$) can be determined through the equation:         
\begin{equation}
	\gamma = -(dln\omega/dlnV)_{P=0} = (B_0/\omega_0)(d\omega/dP)_{P=0}
\end{equation}
Here, $\omega_0$ and $\omega$ represent the Raman frequency under ambient and HP conditions, respectively. $B_0$ stands for the bulk modulus specific to a particular phase. The Gruneisen parameters ($\gamma$) for the cubic and HP tetragonal phases fall within the range of 0.82-1.07 and 1.07-1.24, respectively. Additionally, the Gruneisen parameters ($\gamma$) for the HP orthorhombic Fddd and Cmcm phases vary between 0.10-1.08 and 0.11-1.09, respectively, as detailed in Tables IV, and V.

Moreno et al. reported that the post-spinel orthorhombic structure of $CaTi_2O_4$ in the Cmcm space group exhibits 18-24 Raman-active modes at higher pressures \cite{Moreno2011}. 
As mentioned in our HP XRD study, the proportion of the Cmcm space group increases with pressure, reaching its maximum at the highest pressure level. Therefore, we expect the occurrence of phonon modes associated with the Cmcm-orthorhombic phase at higher pressures in the Raman spectrum. However, in our HP Raman study, no Raman modes were observed beyond 23 GPa, as previously discussed, even though the XRD peaks remain strong. This absence of phonon modes, along with the minimum in FWHM, provides strong evidence for the metallization of the sample under HP conditions. 

To further investigate this, we performed DFT analysis to examine the pressure-dependent metallization of the system. Our calculations employed a similar magnetic configuration for the ambient pressure structure as described by Ghosh et al.\cite{Ghosh2018}. The three phases [cubic, tetragonal, and orthorhombic (Fddd)] exhibit antiferromagnetic ordering with Co atoms in the tetrahedra and octahedra contributing to up-spin and down-spin states, respectively. In each phase, the Co$^{2+}$ ions having three unpaired electrons occupy both tetrahedral and octahedral sites with a high-spin configuration\cite{Ghosh2018}. These unpaired electrons contribute to a total magnetic moment of 3 $\mu_B$. Our calculations yield the following magnetic moments per Co site for the phase: (a) cubic (Fd$\bar{3}$m): $\mu_{tet}$ =2.696 $\mu_B$ and $\mu_{oct}$ = -2.685$\mu_B$, (b) the tetragonal (I4$_1$/amd): $\mu_{tet}$ = 2.664 $\mu_B$ and $\mu_{oct}$ = -2.681 $\mu_B$, and (c) orthorhombic (Fddd): $\mu_{tet}$ = 2.649 $\mu_B$ and $\mu_{oct}$ = -2.672 $\mu_B$. These results are in good agreement with previous theoretical studies \cite{Ghosh2018}. Notably, the other orthorhombic (Cmcm) phase exhibits a net ferrimagnetic ground state with Co magnetic moments $\mu_{tet}$ = 2.649 $\mu_B$ and $\mu_{oct}$ = -0.592 $\mu_B$. Our calculations reveal that the Co$^{2+}$ ions at the octahedral sites undergo a high-spin to low-spin transition in the Cmcm-orthorhombic phase. This transition at the octahedral sites is directly linked to the unit cell and octahedral volume collapse at the Cmcm-orthorhombic phase boundary. Figure 9 shows the density of states (DOS) for the four phases in their antiferromagnetic/ferrimagnetic configurations. It has been found that in the cubic (Fd$\bar{3}$m), tetragonal (I4$_1$/amd), and orthorhombic (Fddd) phases, the energy gap is close to 0.53 eV, 0.43 eV, and 0.28 eV, respectively, whereas the orthorhombic (Cmcm) phase exhibits a metallic character.    
 
Typically, binary and ternary transition metal oxides exhibit insulating behaviour under ambient pressure and temperature conditions \cite{Rahman2017,Rozenberg2006,Okada2008,Furubayashi1997}. This insulating nature arises from strong electronic correlation phenomena inherent in the d-bands, classifying them as Mott insulators \cite{Mott1990}. Mott insulators are characterized by substantial optical d-d gaps, denoted as U, which separate the upper empty d-band from the lower filled one (Mott-Hubbard gap). Alternatively, they can be characterized by a p-d gap, represented by $\Delta$, which separates the upper p-band from the filled d-band (charge-transfer gap). At ambient conditions, Mott insulators often display large magnetic moments, transitioning into an antiferromagnetic state below the Neel temperature ($T_N$). The introduction of chemical doping or the application of pressure to Mott insulators has the potential to close the Mott-Hubbard or charge-transfer gap. This process disrupts the Mott-Hubbard d-d correlation, leading to an Insulator-to-Metal Transition (IMT) accompanied by a collapse of the magnetic moment—a phenomenon known as the Mott transition\cite{Pasternak1999}. It is evident that in Mott insulators, a significant volume reduction during a structural transition must be attributed to the occurrence of the Mott transition \cite{Rozenberg2002,Greenberg2013,Arielly2011}. In an extensive HP investigation of $Zn_{0.2}Mg_{0.8}Fe_{2}O_{4}$, Rahman et al. \cite{Rahman2017} 
observed a substantial $7.5\%$ change in the unit cell volume during the cubic to orthorhombic phase transition. Concurrently, Raman modes vanished at higher pressure, leading them to propose a Mott Insulator-to-Metal Transition (IMT) around 21 GPa. Pasternak et al. \cite{Pasternak1999} reported a first-order IMT induced by pressure in the Mott insulator $Fe_2O_3$, characterized by the breakdown of d-d correlation and a collapse of magnetism. Greenberg et al. \cite{Greenberg2013} observed a Mott transition in $CaFe_2O_4$ around 50 GPa, associated with an isostructural transition and a sharp $12\%$ decrease in Fe polyhedral volume due to the closure of the Hubbard gap. This transition was accompanied by a high-spin to low-spin transition, leading to the disappearance of most Raman modes and the persistence of a few low-intensity modes. In a recent theoretical study, Leonov et al. \cite{Leonov2020} identified a pressure-induced Mott insulator-to-metal transition in CoO, MnO, FeO, and NiO. This transition was marked by the simultaneous collapse of local magnetic moments and lattice volume, achieved through a high-spin to low-spin crossover and band-gap closure driven by a change in crystal structure. They also propose the occurrence of quantum critical charge and spin fluctuations near the pressure-driven Mott IMT in correlated systems with spin-state active ions. Consequently, high pressure not only alters crystal symmetry but also induces Mott transition breakdown, coupled with a spin crossover from high-spin to low-spin, leading to the collapse of local magnetic moments and volume reduction. These phenomena are common in many other cobaltites and ferrites, all demonstrating pressure-driven structural phase transitions characterized by the contraction of metal-oxygen polyhedra and a significant reduction in unit cell volume at the transition.

{In our HP XRD investigation, a structural transition is observed at approximately 17.3 GPa, shifting from a tetragonal phase to two coexisting orthorhombic phases}. During the tetragonal to Cmcm-orthorhombic phase transition, a pronounced reduction in unit cell volume of approximately $17.5\%$ is noted at the transition pressure, indicating a first-order phase transition with a significant collapse in volume. Additionally, the volume of $Co/TiO_6$ octahedra experienced a collapse of around $16\%$ at the phase transition point (refer to Figure 4(c)). Our DFT calculations reveal that Co ions at the octahedral sites undergo a high-spin to low-spin transition in the high-pressure Cmcm-orthorhombic phase, which also exhibits metallic behaviour. We also observed an increase in the distortion index (DI) of $Co/TiO_6$ octahedra at the phase boundary (see Figure 6(b), (c)). This pressure-induced rise in DI results in an increase in the crystal field, contributing to a substantial decrease in the magnetic moment of the metal ions in the B site \cite{Rahman2017}. As previously discussed, the observed minimum in the FWHM of the $T_{2g}(2)$ and $A_{1g}$ Raman modes suggests a possible electronic transition near 17.3 GPa. This transition is likely attributed to the pressure-induced closure of the charge-transfer gap, resulting in metallization accompanied by a magnetic collapse. The significant reductions in the unit cell and octahedral volume, along with a high-spin to low-spin transition associated with a spin crossover, breakdown the Mott-Hubbard d-d correlation, leading to the metallic behaviour of CTO-Sp above 17.3 GPa, accompanied by the collapse of local magnetic moments.  A possible electronic transition due to the pressure-induced bandgap closure, the absence of all Raman modes, combined with the non-appearance of Raman modes responsible for the Cmcm-orthorhombic phase at HP, further supports the notion of a pressure-induced metallization in our sample.
                                                                      
\section{Conclusion}
 We have investigated the pressure-induced structural, vibrational, and electronic properties, as well as possible magnetic transition, in the inverse spinel $Co_2TiO_4$ using X-ray diffraction, Raman spectroscopy, and first-principles DFT calculations. Our experiments reveal two structural phase transitions. The first transition, occurring around 7.3 GPa, involves a cubic-to-tetragonal transformation. The second phase transition takes place at approximately 17.3 GPa, {Involving a shift from the tetragonal phase to two coexisting orthorhombic phases with space groups Fddd and Cmcm.} The tetragonal to Cmcm-orthorhombic phase transition is characterized by a first-order phase transition accompanied by a substantial volume collapse (17.5\% for the unit cell volume and 16\% for $Co/TiO_6$ octahedral volume). The proportion of the Cmcm-orthorhombic phase increases with pressure, reaching its peak at the maximum pressure. The macroscopic bulk modulus for the different phases of CTO-Sp can be estimated as the average polyhedral bulk modulus of the respective phases. The significant reductions in unit cell and octahedral volume, coupled with the high-spin to low-spin transition, result in a pressure-induced metallization, leading to the collapse of local magnetic moments in the Cmcm-orthorhombic phase. The emergence of a new peak and changes in the slope of $T_{2g}(2)$ and $A_{1g}$ modes around 17.3 GPa, as observed in Raman spectroscopy, align with structural reorientation, consistent with XRD measurements. The absence of Raman modes, coupled with the non-appearance of modes linked to the Cmcm-orthorhombic phase beyond 23 GPa, and a possible electronic transition due to the pressure-driven closure of the charge-transfer gap, further supports the idea of a pressure-induced metallization in CTO-Sp. Our results suggest that CTO-Sp undergoes a possible Mott IMT accompanied by a possible magnetic transition around 17.3 GPa, which holds significant implications for understanding the quantum criticality of the Mott transition. Further theoretical studies and experiments on cobaltites are essential to comprehensively characterize the pressure-driven metallization process.
                                 
\begin{acknowledgments}
 The authors gratefully acknowledge the financial support from the Department of Science and Technology, Government of India, to visit the XPRESS beamline in the ELETTRA Synchrotron light source under the Indo-Italian Executive Programme of Scientific and Technological Cooperation. MS and SC gratefully acknowledge the CSIR, Government of India, for the financial support to carry out the PhD work.	 
\end{acknowledgments}
\noindent
{\bf {Author Declarations}} 
\noindent All authors have an equal contribution. All authors reviewed the manuscript.\\

\pagebreak
	
	\begin{table}
		\centering
		\small
		\setlength{\tabcolsep}{6pt} 
		\caption{\textbf{Rietveld-refined structural parameters for the XRD pattern of CTO-Sp are listed for both the ambient cubic phase and the tetragonal phase at 9.6 GPa. The goodness-of-fit is characterized by $R_{wp} = 4.2\%$, $R_{p} = 1.6\%$ (for cubic) and $R_{wp} = 4.1\%$, $R_{p} = 2.2\%$ (for tetragonal)}}
		\begin{tabular}{p{0.9cm} p{0.9cm} p{1cm} p{0.9cm} p{0.9cm} p{1.2cm} p{0.9cm} p{0.9cm} p{0.9cm} p{0.9cm} p{0.9cm}}
			\hline\hline
			& \multicolumn{4}{c}{Cubic ($Fd\bar{3}m$)} & \multicolumn{5}{c}{Tetragonal ($I4_1/amd$)} \\
			\hline
			Atoms & Wyck. & Occu. & $x/a$ & $y/b$ & $z/c$ & Wyck. & Occu. & $x/a$ & $y/b$ & $z/c$ \\
			\hline  
			Co1 & 8b & 1 & 0.3750 & 0.3750 & 0.3750 & 4a & 1 & 0.0000 & 0.7500 & 0.1250 \\ 
			Co2 & 16c & 0.5 & 0.0000 & 0.0000 & 0.0000 & 8c & 0.5 & 0.0000 & 0.5000 & 0.5000 \\
			Ti & 16c & 0.5 & 0.0000 & 0.0000 & 0.0000 & 8c & 0.5 & 0.0000 & 0.5000 & 0.5000 \\
			O & 32e & 1 & 0.2430 & 0.2430 & 0.2430 & 16h & 1 & 0.0000 & 0.0196 & 0.2606 \\
			\hline
			a=8.4473(7)$\AA$, V=602.788(13)$\AA^3$ & & & & & & a=5.8845(7)$\AA$, c=8.3374(5)$\AA$ V=288.711(1)$\AA^3$ \\
			\hline\hline
		\end{tabular}
	\end{table}

\begin{table}
	\centering
	\small
	\setlength{\tabcolsep}{6pt} 
	\caption{\textbf{Rietveld-refined structural parameters for the XRD pattern of CTO-Sp are listed for both the $Fddd$ and $Cmcm$ space group in the orthorhombic phase at 19.1 GPa. The goodness-of-fit is characterized by $R_{wp} = 3.7\%$ and $R_{p} = 1.8\%$}}
	\begin{tabular}{p{0.9cm} p{0.9cm} p{1cm} p{0.9cm} p{0.9cm} p{1.2cm} p{0.9cm} p{0.9cm} p{0.9cm} p{0.9cm} p{0.9cm} p{0.9cm}}
		\hline\hline
		& \multicolumn{4}{c}{Orthorhombic: $Fddd$} & \multicolumn{5}{c}{Orthorhombic: $Cmcm$} \\
		\hline
		Atoms & Wyck. & Occu. & $x/a$ & $y/b$ & $z/c$ & Atom & Wyck. & Occu. & $x/a$ & $y/b$ & $z/c$ \\
		\hline  
		Co1 & 8a & 1 & 0.1250 & 0.1250 & 0.1250 & Co1 & 4c & 1 & 0.0000 & 0.3809 & 0.2500 \\ 
		Co2 & 16d & 0.5 & 0.5000 & 0.5000 & 0.5000 & Co2 & 8f & 0.5 & 0.0000 & 0.1331 & 0.0753 \\
		Ti & 16d & 0.5 & 0.5000 & 0.5000 & 0.5000 & Ti & 8f & 0.5 & 0.0000 & 0.1331 & 0.0753 \\
		O & 32h & 1 & 0.2610 & 0.2650 & 0.2590 & O1 & 4c & 1 & 0.0000 & 0.0581 & 0.2500 \\
		  &&&&&&								 O2 & 8f & 1 & 0.0000 & 0.2421 & 0.6028 \\
		  &&&&&&								 O3 & 4b & 1 & 0.0000 & 0.5000 & 0.0000\\										
		\hline
		a=8.2650(2)$\AA$, b=8.3321(7)$\AA$, c=8.1670(6)$\AA$ V=562.41(6)$\AA^3$ & & & & & & a=2.7521(7)$\AA$, b=8.8650(2)$\AA$, c=9.5185(8)$\AA$ V=232.220(7)$\AA^3$ \\
		\hline\hline
	\end{tabular}
\end{table}

\begin{table}
	\centering
	\small
	\setlength{\tabcolsep}{6pt}
	\caption{\textbf{Macroscopic and polyhedral values of $B_{0}$ and $B^{'}_{0}$ for various phases of CTO-Sp. $\bar{B_{0}}$ represents the average polyhedral bulk modulus, as defined in the text. The unit of bulk modulus is GPa.}}
	\begin{tabular}{p{2cm} p{2.2cm} p{3cm} p{3cm} p{3cm}}
		\hline\hline
		& Cubic: $Fd\bar{3}m$ & Tetragonal: $I4_1/amd$ & Orthorhombic: $Fddd$ & Orthorhombic: $Cmcm$\\
		\hline
		 $B_0$ & 203(2) & 235(7) & 217(2) & 230(2) \\
		 $B^{'}_0$ & 3.8(0.1) & 10(2) & 2.7(0.1) & 3.7(0.3)\\
		 $(B_0)_{tet}$ & 204(2) & 260(1) & 222(1) & ...\\
		 $(B^{'}_0)_{tet}$ & 3.5 (0.6) & 6.6(0.3) & 0.6(0.05) & ...\\
		 $(B_0)_{oct}$ & 202(1) & 236(7) & 206(4) & 231(4) \\
		 $(B^{'}_0)_{oct}$ & 3.8(0.3) & 10(1.5) & 1.5(0.3) & 3.91(0.4)\\
		 $(B_{0})_{Co^{2+}O_6}$ & ... & ... & ... & 232(2)\\
		 $(B^{'}_{0})_{Co^{2+}O_6}$ & ... & ... & ... & 3.8(0.15)\\
		 $\bar{B_0}$ & 203(3) & 248(8) & 214(5) & 231.5(6)\\
		 \hline\hline
		 \end{tabular}
	\end{table}

\begin{table}
	\centering
	\setlength{\tabcolsep}{6pt}
	\caption {\textbf{Raman modes, their pressure coefficients $d\omega/dP$, and calculated mode Gruneisen parameters ($\gamma$) for the cubic and tetragonal phases of CTO-Sp.}}
	\begin{tabular}{p{4.5cm}| p{4cm} p{1.5cm}|p{4cm} p{1cm}}
		& Cubic: $Fd\bar{3}m$ &  & Tetragonal: $I4_1/amd$ \\
		\hline\hline
		Raman modes $\omega_0$ ($\text{cm}^{-1}$) & $d\omega/dP$$(cm^{-1}/GPa)$ & $\gamma$ & $d\omega/dP$$(cm^{-1}/GPa)$ & $\gamma$ \\
		\hline
		174 ($T_{2g}(1)$) & 0.74 & 0.82 & .... & ....\\
		309.4 ($E_g$) & 1.37 & 0.89 & 1.81 & 1.24\\
		531.6 ($T_{2g}(2)$) & 2.39 & 0.87 & 2.39 & 1.07\\
		701.8 ($A_{1g}$) & 3.77 & 1.07 & 3.77 & 1.20\\
		\hline\hline
	\end{tabular}
\end{table}

\begin{table}
	\centering
	\setlength{\tabcolsep}{6pt}
	\caption{\textbf{Raman modes, their pressure coefficients $d\omega/dP$, and calculated mode Gruneisen parameters ($\gamma$) for the different space groups of the post-spinel orthorhombic phase of CTO-Sp.}}
	\begin{tabular}{p{4.5cm}| p{4cm} p{1.5cm}|p{4cm} p{1cm}}
		& Orthorhombic: $Fddd$ &  & Orthorhombic: $Cmcm$ \\
		\hline\hline
		Raman modes $\omega_0$ ($\text{cm}^{-1}$) & $d\omega/dP$$(cm^{-1}/GPa)$ & $\gamma$ & $d\omega/dP$$(cm^{-1}/GPa)$ & $\gamma$ \\
		\hline
		174 ($T_{2g}(1)$) & .... & .... & .... & ....\\
		309.4 ($E_g$) & .... & .... & .... & ....\\
		576 ($T_{2g}(2)$) & 0.28 & 0.10 & 0.28 & 0.11 \\
		764 ($A_{1g}$) & 3.09 & 0.90 & 3.09 & 0.90\\
		690 (new mode, M1) & 3.32 & 1.08 & 3.32 & 1.09\\
		\hline\hline
	\end{tabular}
\end{table}
\pagebreak

  \begin{figure}[htb]
 	\begin{center}
 		\includegraphics[width=\linewidth]{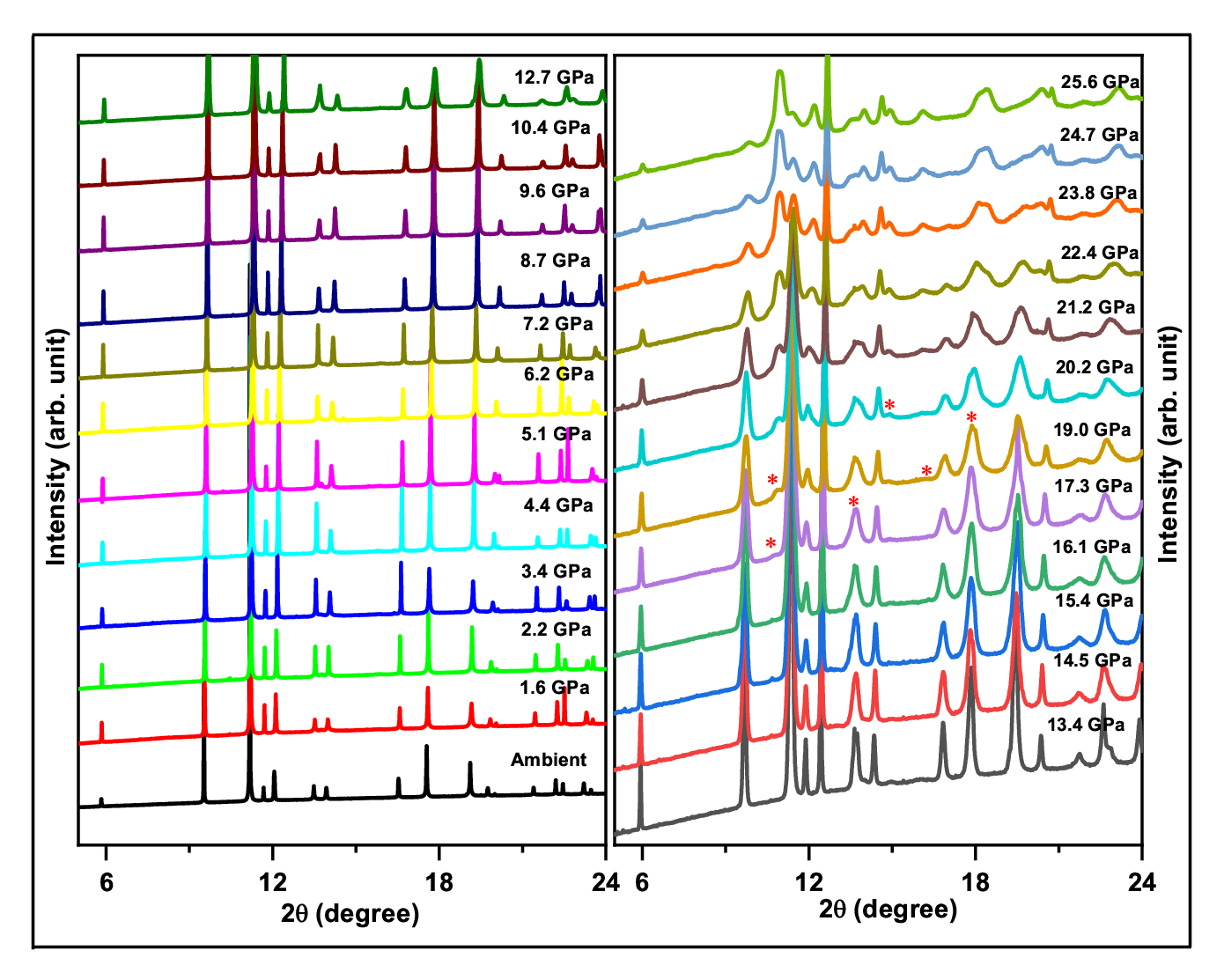}
 		\caption{X-ray powder diffraction patterns of $Co_2TiO_4$ at selected pressures.}
 	\end{center}
 \end{figure}

\begin{figure}[htb]
	\begin{center}
		\includegraphics[width=0.6\linewidth]{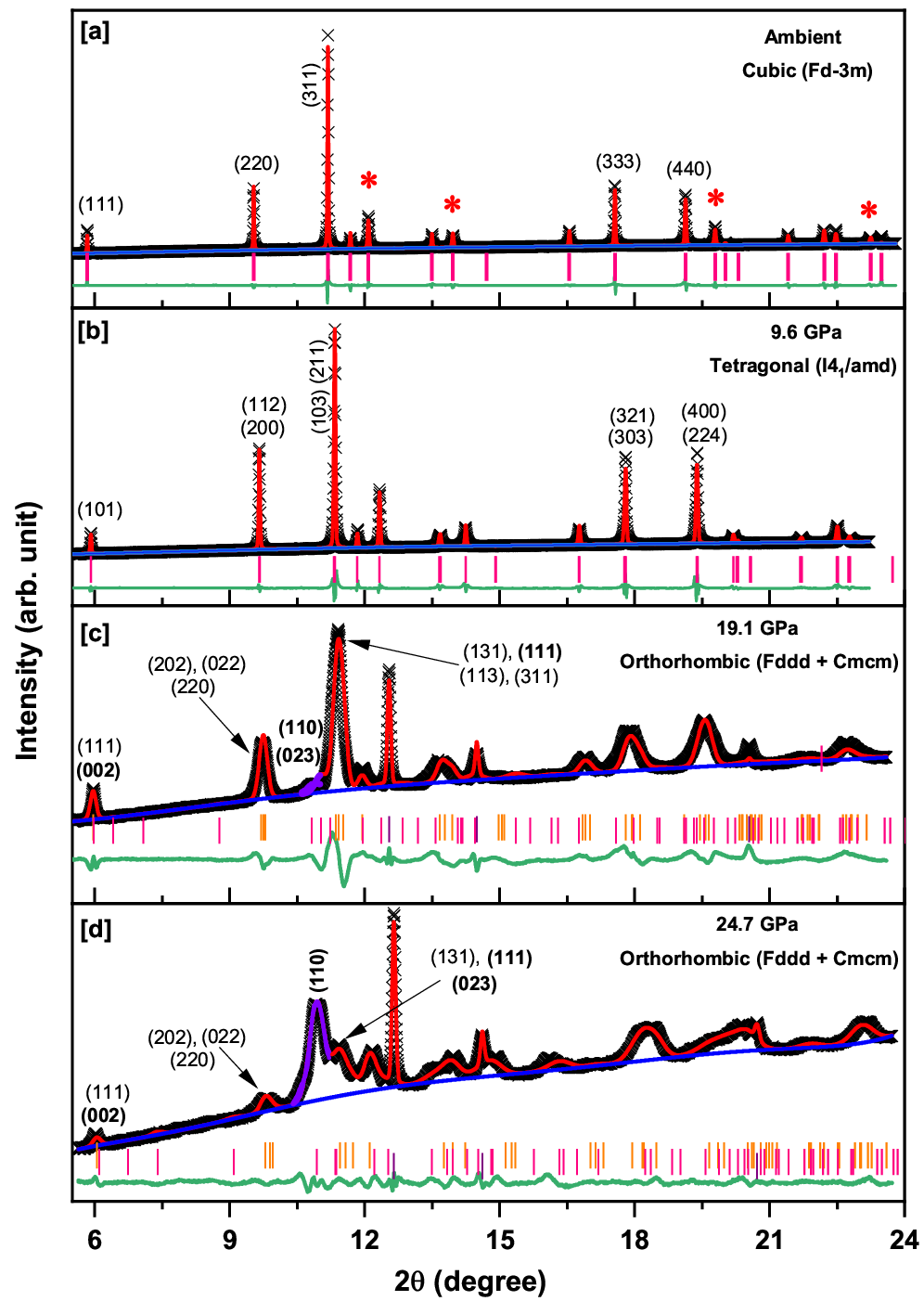}
		\caption{Rietveld refinement of X-ray powder diffraction patterns for $Co_2TiO_4$ at [a] ambient pressure (cubic phase, space group: Fd$\bar{3}$m) [b] 9.6 GPa (tetragonal phase, space group: $I{4_1}/amd$) [c] 19.1 GPa, and [d] 24.7 GPa (orthorhombic phase, space groups: Fddd + Cmcm). The black crosses represent observed data points, while the red and violet lines over the data points depict the fit to the data. The intensity of the peak corresponding to the Cmcm space group increases with pressure, as shown by the violet line. The blue line indicates the background, and the difference between observed and fitted data is illustrated by the green line. Bragg positions are marked by vertical bars. In the orthorhombic phase, the orange and pink vertical bars represent the Bragg positions of the Fddd and Cmcm space groups, respectively. Diffraction peaks of the pressure marker are indicated by red star markers. {Selected major peaks are labeled with their (hkl) values, and the peaks from the Cmcm phase are shown in bold.}}
	\end{center}
\end{figure}

\begin{figure}[htb]
	\begin{center}
		\includegraphics[width=\linewidth]{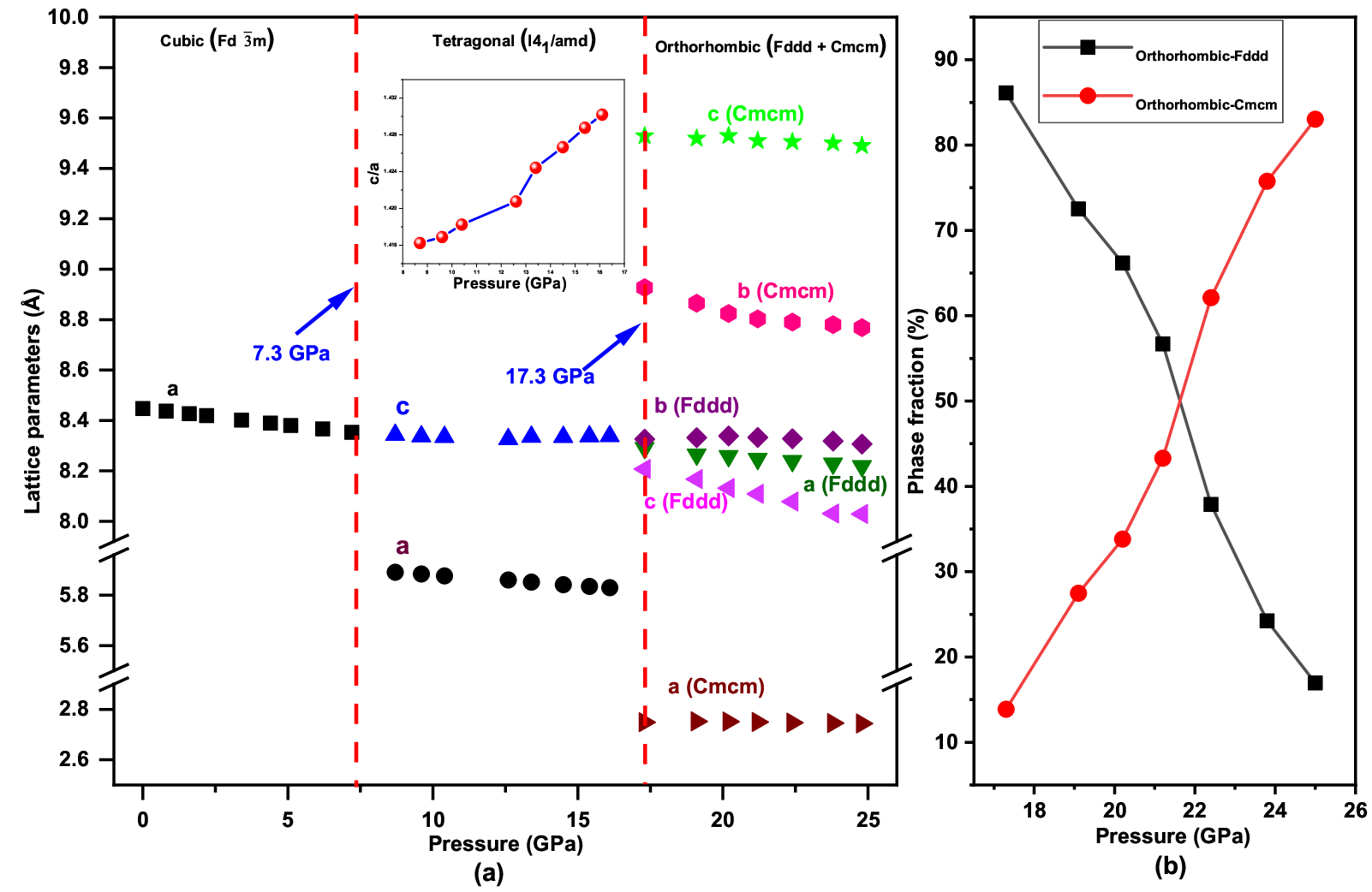}
		\caption{(a) Lattice parameter of the unit cells for the three different phases. In the tetragonal phase of $Co_2TiO_4$, the tetragonality (c/a) as a function of pressure is illustrated in the inset. [b] Pressure-dependent phase fraction percentage for the two different space groups in the orthorhombic phase.}
	\end{center}
\end{figure}

\begin{figure}[htb]
	\begin{center}
		\includegraphics[width=\linewidth]{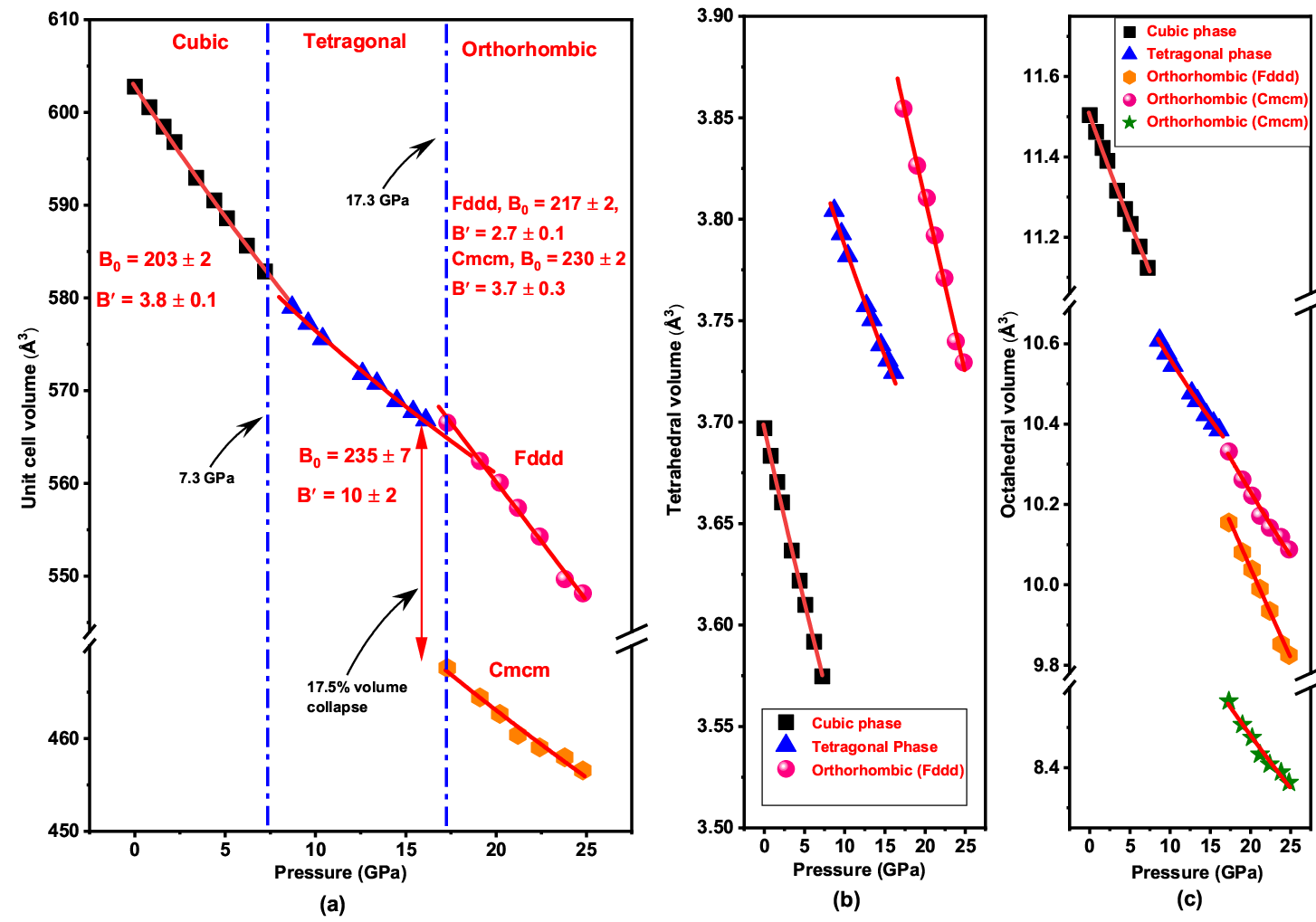}
		\caption{Evolution of pressure: (a) unit cell volume of polycrystalline CTO-Sp across three phases, (b) tetrahedral volume for various phases, (c) octahedral volume for the three phases. Solid lines through the data points represent the fitting using the Birch-Murnaghan equation of state (BM-EoS).}
	\end{center}
\end{figure}

\begin{figure}[htbp]
	\begin{center}
		\includegraphics[width=\linewidth]{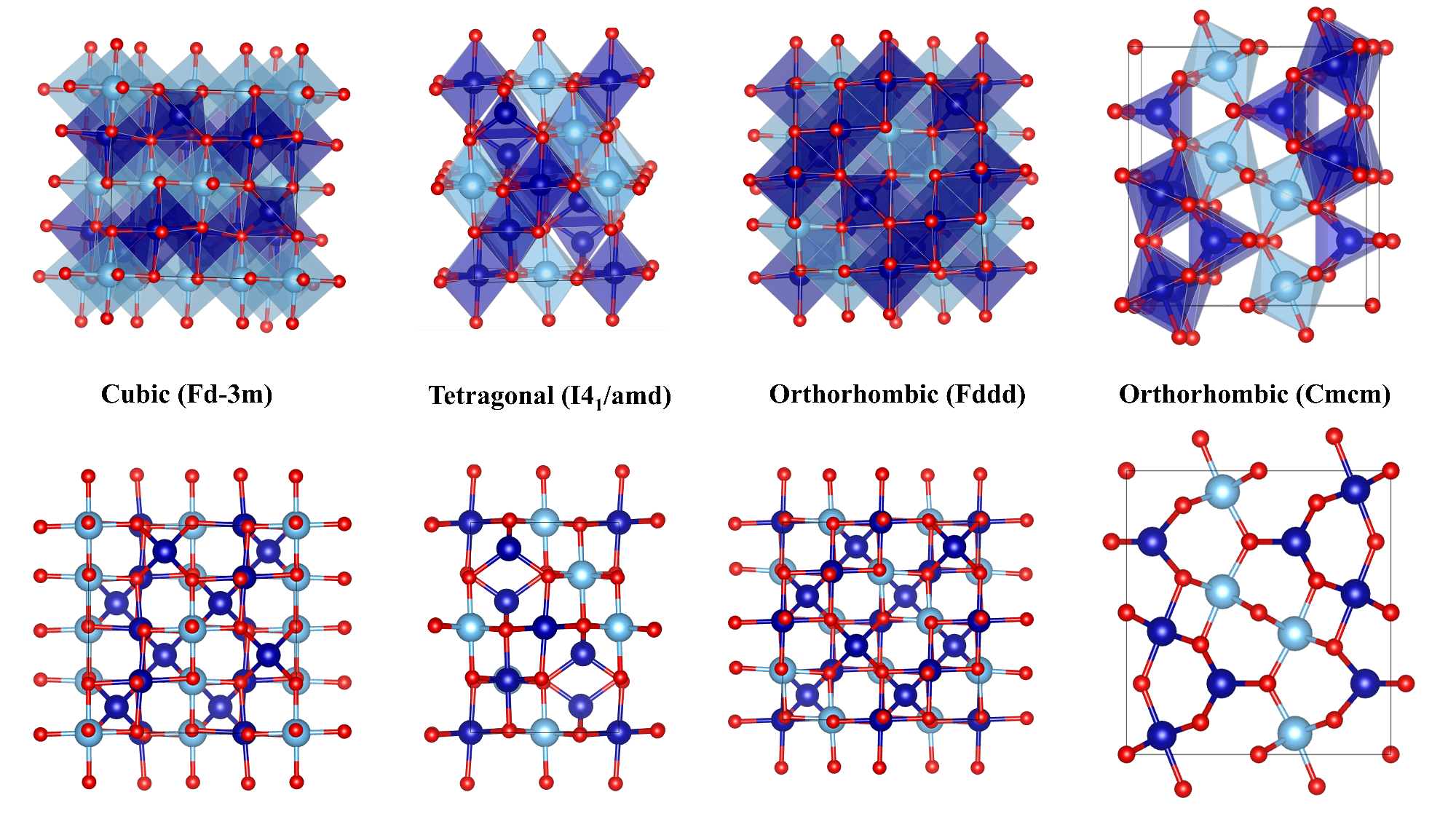}
		\caption{The optimized crystal structures for the different phases. The initial lattice parameters and the atom positions are obtained from our Rietveld refined XRD data.}
	\end{center}
\end{figure}

\begin{figure}[htb]
	\begin{center}
		\includegraphics[width=\linewidth]{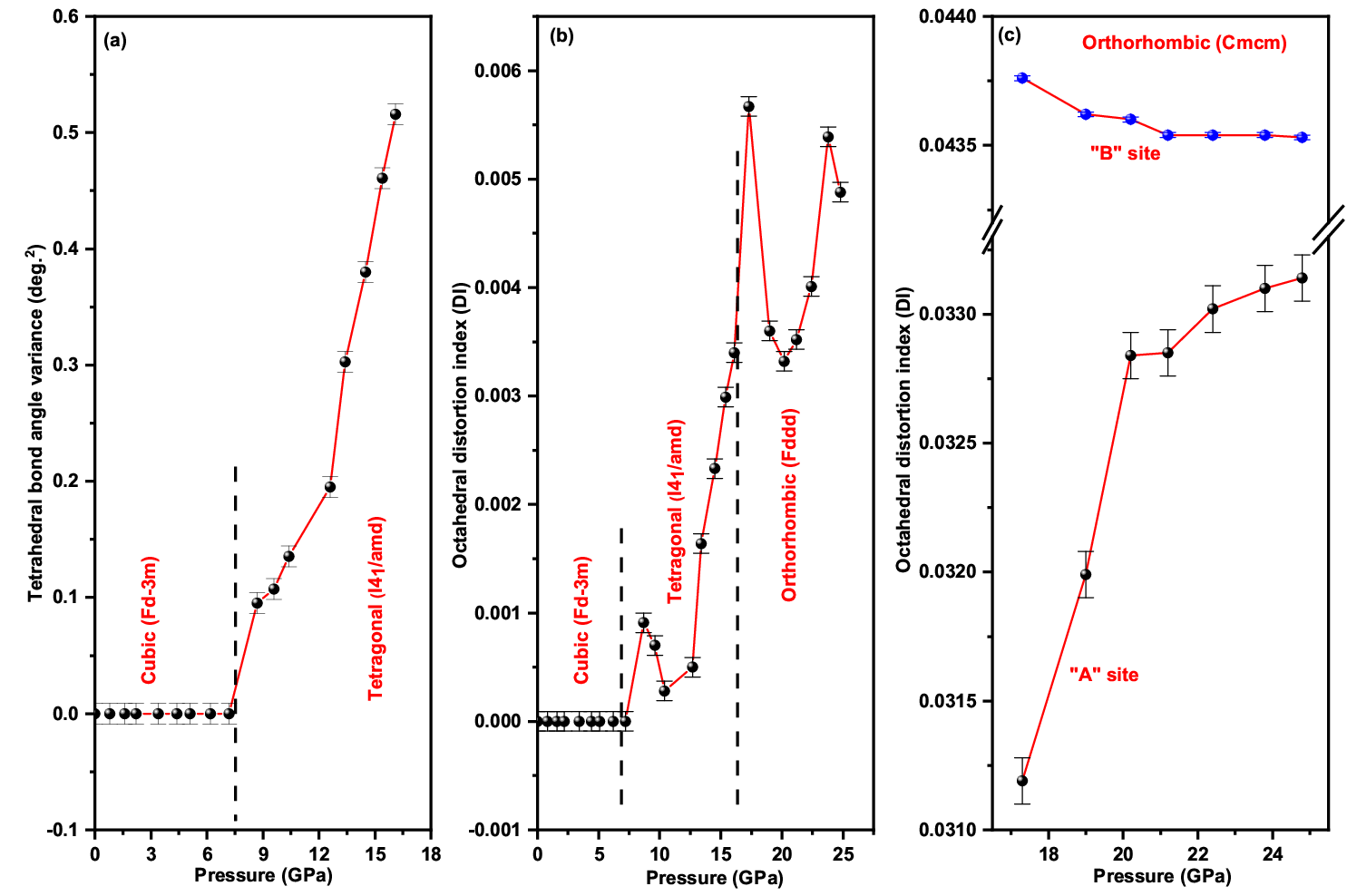}
		\caption{pressure-dependent variations: (a) tetrahedral bond angle variance in both cubic and tetragonal phases, (b) octahedral distortion index (DI) in cubic ($Fd\bar{3}m$), Tetragonal ($I4_1/amd$), and orthorhombic ($Fddd$) phases, and (c) octahedral DI specifically for the orthorhombic-Cmcm phase. Connecting lines between the data points serve as visual guides.}
	\end{center}
\end{figure}

\begin{figure}[htb]
	\begin{center}
		\includegraphics[width=0.7\linewidth]{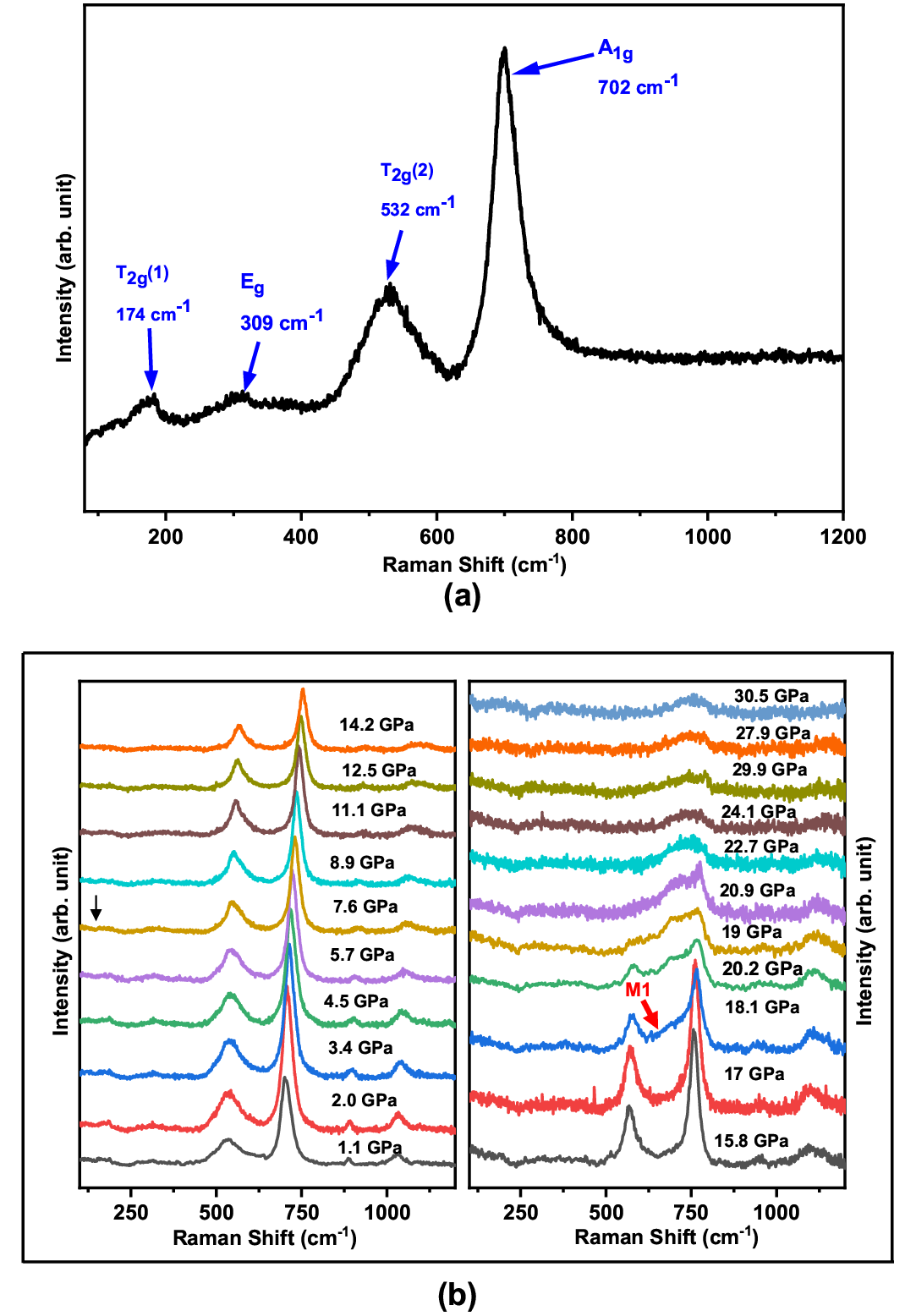}
		\caption{(a) Raman spectrum of $Co_2TiO_4$ at ambient conditions, obtained with a 532 nm laser as the excitation source. All modes are labelled in the figure. (b) Pressure evolution of Raman spectra at selected pressure points. The emergence of a new Raman mode in the orthorhombic phase is indicated by M1.}
	\end{center}
\end{figure}

\begin{figure}[htb]
	\begin{center}
		\includegraphics[width=\linewidth]{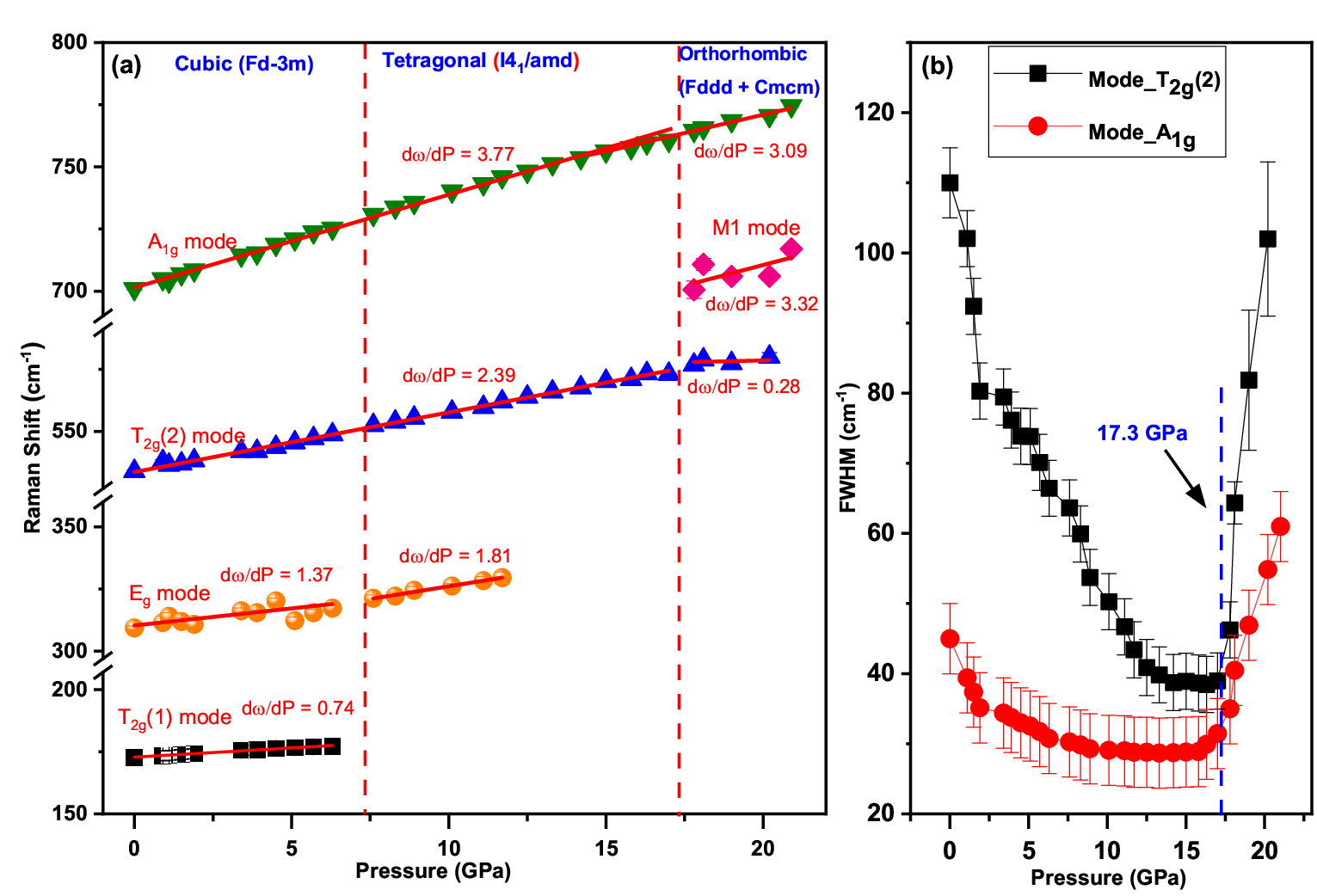}
		\caption{(a) Raman mode values observed in polycrystalline $Co_2TiO_4$ as a function of pressure. Slope discontinuities along with the disappearance of phonon modes observed at 7.3 GPa and 17.3 GPa. A new mode M1 observed in the HP orthorhombic phase. {Solid symbols represent the frequency of Raman modes at different Pressures. Solid lines indicate linear fit to the data.} (b) Pressure-dependent Full Width at Half Maximum (FWHM) of the $T_{2g}(2)$ and $A_{1g}$ modes, revealing a minimum at approximately 17.3 GPa.}
	\end{center}
\end{figure}
\begin{figure}[htbp]
	\begin{center}
		\includegraphics[width=\linewidth]{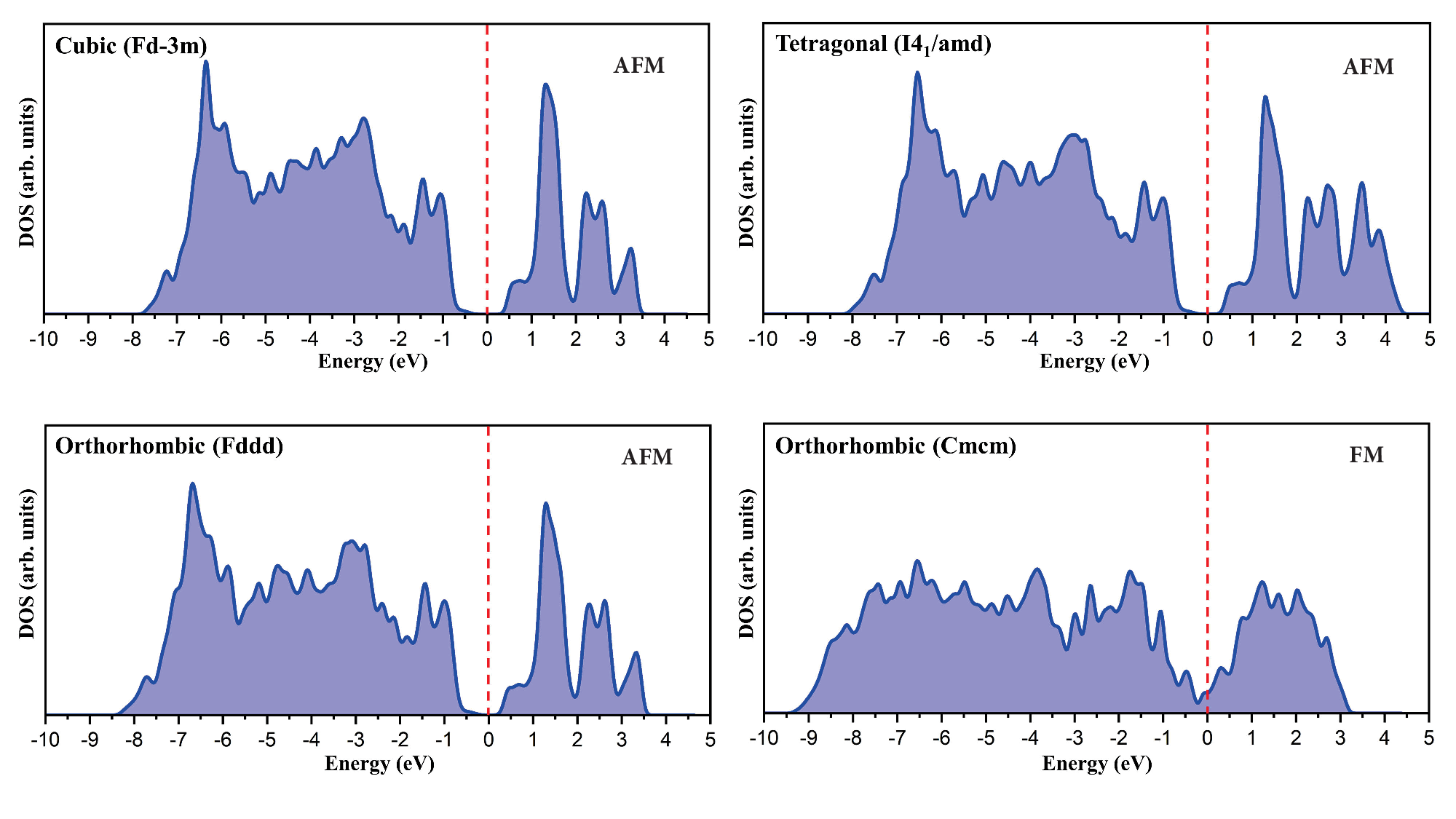}
		\caption{Total electronic density of states for the four different phases in their antiferromagnetic (AFM)/ferromagnetic (FM) configurations with U$_{eff}$ = 4.0 eV for Co ions. {Fermi level has been marked with red dotted lines.}}
	\end{center}
\end{figure}
\end{document}